\documentclass[aip,jcp,preprint]{revtex4-1}
\usepackage[utf8]{inputenc}

\usepackage{amsmath}
\usepackage{amsfonts}
\usepackage{amssymb}
\usepackage{braket}
\usepackage{bm}

\usepackage{graphics}
\usepackage{xcolor}

\usepackage{tikz}
\usetikzlibrary{arrows}
\usetikzlibrary{decorations.pathmorphing}
\usetikzlibrary{decorations.markings}
\usetikzlibrary{decorations.pathreplacing}

\usepackage{natbib}

\newcommand{\Tr}{\text{Tr}}
\newcommand{\Real}{\text{Re}\,}
\newcommand{\Imag}{\text{Im}\,}

\newcommand{\super}[1]{\mathcal{#1}}
\newcommand{\commute}[2]{\left[#1, #2\right]}
\newcommand{\field}{\varepsilon}
\newcommand{\ft}{\tilde}

\begin{document}
\title{Considerations regarding one-photon phase control}
\date{\today}
\author{Cyrille Lavigne}
\author{Paul Brumer}
\affiliation{Chemical Physics Theory Group, Department of Chemistry, and Center for Quantum Information and Quantum Control, University of Toronto, Toronto, Ontario, M5S 3H6, Canada}

\begin{abstract}
    The dynamics of a system interacting with an ultrashort pulse is
    known to depend on the phase content of said pulse.  For linear
    absorption, phase control is possible over time-varying quantities,
    such as the population of metastable states, but not over
    time-independent quantities, such as the population of steady states.
    We derive here a strict upper bound for phase control that
    interpolates between these two cases --- the bound quantifies the
    approach to the steady state and resulting loss of one-photon phase
    control based on physical timescales.  Significantly, this bound is
    violated by a number of numerical and experimental investigations.  A
    careful analysis of the physical conditions underlying this result
    exposes multiphoton effects as a mechanism for these experiments.
\end{abstract}
\maketitle

A quantum mechanical system interacting with light is sensitive to the
phase of the light, which forms the physical basis for phase
control.\cite{shapiro_quantum_2012}  Specifically, the response of a
quantum mechanical system to monochromatic radiation is dependent on
the intensity and the frequency of radiation.  When the exciting field
is multichromatic (i.e. it includes more than one frequency
component), the system's response becomes dependent not only on the
intensity of each frequency component but also on the phase
differences between these components.  The latter is the source of
phase control.  In particular, coherent control, the use of rationally
designed and physically motivated control procedures to exploit
molecular interference, has been successfully demonstrated in a
variety of
contexts.\cite{zhu_coherent_1995,hache_observation_1997,stievater_rabi_2001}

Phase interference arises from the wave nature of light, of the system
under control or a combination of both.\cite{franco_quantum_2010}
Properties of interference have been used to devise strict conditions
on the possibility of coherent control in a variety of scenarios.
\cite{
  brumer_one_1989,spanner_communication:_2010,
  mukamel_coherent-control_2013,am-shallem_scaling_2014,
  shapiro_quantum_1995,wu_quantum_2008,pachon_mechanisms_2013}
These conditions have an important role in informing the
interpretation of coherent control experiments.  

Control from one-photon absorption is particularly interesting because
the linear absorption probability is independent of the phase of the
absorbed light.\cite{brumer_one_1989}  Thus, phase control in the
linear regime is never due to a phase-dependent change in the
absorption probability, as may be the case in multiphoton
absorption.\cite{brenner_two-photon_2013,lavigne_ultrashort_2019}
This phase insensitive absorption probability greatly simplifies the
interpretation of one-photon phase control.  Consider, for example,
the \textit{cis-trans} isomerization of
retinal.\cite{birge_energy_1983}  In a linear intensity experiment, if
modifying the phase of the exciting light leads to a change in the
isomerization yield, it follows immediately that direct phase control
over the isomerization itself has been achieved.  In contrast,
multiphoton absorption is phase-dependent\cite{faisal_theory_1987} ---
a change in the isomerization yield in a multiphoton experiment might
reflect phase control over the
absorption with no
change in the isomerization
dynamics.\cite{ogilvie_fourier_2005,hoki_mechanisms_2005,brenner_two-photon_2013,lavigne_ultrashort_2019}

Experimental and numerical demonstrations of one-photon phase control
lead to the development of formal conditions under which such control
is possible.  In the control of molecular photodissociation, it was
shown that the cross-section of a photochemical reaction (a
steady-state quantity) triggered by a one-photon excitation is
independent of the phase of the excitation.\cite{brumer_one_1989}  It
was proven later that the phase of an exciting field (in a one-photon
experiment where the initial state is time-independent) can only
control quantities exhibiting time-dependence (e.g. the population of
metastable states); time-independent properties (e.g. the population
of steady-states) are insensitive to the phase of
light.\cite{spanner_communication:_2010,mukamel_coherent-control_2013}
That is, one-photon phase control is always time-dependent.

The dynamical nature of control is easily observed in small quantum
systems, where controllable properties are highly oscillatory or
rapidly decaying.\cite{
  christopher_overlapping_2005,christopher_quantum_2006,
  brinks_visualizing_2010,brinks_beating_2011,
  tiwari_laser-pulse-shape_2014,weigel_shaped_2015,
  lavigne_interfering_2017}
To see this, consider a small three-level system coupled to an
electric field $\field(t)$ in the dipole approximation,
\begin{align}
  H(t) = H_0 + \field(t)\mu 
\end{align}
The electric field $\field(t)$ is time-limited such that $\field(t>t_\field) = 0$.
The system is taken to be initially in the state $\ket{g}$, an
eigenstate of $H_0$ with energy $E_g$.  A one-photon excitation 
generates a superposition of two excited eigenstates $\ket{e_1}$ and
$\ket{e_2}$ with energies $E_1$ and $E_2$ with the following wavefunction,\cite{shapiro_principles_2003}
\begin{align}
 \frac{\hbar}{2\pi i} \ket{\psi_1(t>t_\field)} =  \ft\field(\omega_{1g})\mu_{1g} e^{-i E_1 t/\hbar}\ket{e_1} + \ft\field(\omega_{2g})\mu_{2g} e^{-i E_2 t/\hbar}\ket{e_2}
\end{align}
where $\omega_{ij} = (E_i - E_j)/\hbar$, $\mu_{ng} =
\braket{e_n|\mu|g}$ and $\ft\field(\omega)$ is the Fourier transform of $\field(t)$.  The expectation value $\braket{O(t)}$ of an
observable obtained from the photoexcited state, e.g. the population
of a particular isomer in a photoisomerization experiment, is given by
the following expression,
\begin{align}
  \braket{O(t)} = \braket{\psi_1(t)|O|\psi_1(t)} \label{eq:example} &= \frac{4 \pi^2}{\hbar^2} |\ft\field(\omega_{1g})|^2 |\mu_{1g}|^2 O_{11} + \frac{4 \pi^2}{\hbar^2} |\ft\field(\omega_{2g})|^2 |\mu_{2g}|^2 O_{22}\\
  &\quad + \frac{4 \pi^2}{\hbar^2} \ft\field^*(\omega_{1g})\ft\field(\omega_{2g}) \mu^*_{1g} \mu_{2g} O_{12} e^{-i\omega_{21} t/\hbar}  + \text{c.c.}\nonumber
\end{align}
The first two terms depend only on the intensity of the electric field
$|\ft\field(\omega)|^2$ at the two excitation frequencies $\omega_{1g}$
and $\omega_{2g}$ and not on the phase of the electric field.  The
final two terms however are phase-dependent and thus phase
controllable provided that $\omega_{1g} \neq \omega_{2g}$, i.e., that
the states $\ket{e_1}$ and $\ket{e_2}$ are not degenerate.  Such
control however is also time-dependent; the phase-controllable terms
are oscillatory in time with a frequency $\omega_{12}=\omega_{1g} -
\omega_{2g}$ corresponding to the difference between the two
excitation frequencies.

The phase difference between $\ft\field(\omega_{1g})$ and
$\ft\field(\omega_{2g})$ sets the phase of oscillatory terms with
frequency $\omega_{1g} - \omega_{2g}$.  The duration of phase control
is related to the inverse of this frequency $\tau_{12} =
1/(\omega_{1g} - \omega_{2g})$; the average value of $O(t)$ taken over
a time interval $T$ becomes phase-independent when $T \gg \tau_{12}$
and the phase-controllable oscillatory contributions are averaged
out. Phase control, in this case, is ``long-lived'' only in proportion
to the timescale set by $\tau_{12}$; the maximum possible duration
$\tau_{12}$ that can be controlled is itself a function of the
resolution (and thus the duration) of the electric field.
Qualitatively, the maximal duration of control and the duration of the
field are linked.

Results obtained from such a simple model can not be extrapolated to
experimental systems.  In large and in open quantum systems, some
properties, e.g. the population of photoisomers, are dynamical only
over timescales much longer than typical ultrafast
excitations.\cite{lavigne_ultrashort_2019}  Therefore, there are no
\textit{a priori} contradictions\cite{spanner_communication:_2010}
between the dynamical nature of control and the long-lived phase
control reported in some experiments and simulations.\cite{
  herek_quantum_2002,prokhorenko_coherent_2005,
  prokhorenko_coherent_2006,katz_control_2010,
  arango_communication:_2013}
Yet it is also profoundly unsatisfying, from a physical point of view,
that the steady state would be defined only at the $t\rightarrow
\infty$ limit --- what constitutes the ``steady state'' with respect
to control should be related to physical timescales (e.g. $\tau_{12}$
above) of the material system or radiation.

Thus, whether phase effects from one-photon excitations can be stable
or long-lived has remained controversial.  Recently, Kukura et
al. failed to reproduce earlier results using a methodology which, in
principle, should eliminate any non-phase
effects.\cite{liebel_lack_2017}  Steady-state one-photon phase control
has been shown to be impossible for certain classes of open quantum
systems,\cite{am-shallem_scaling_2014} as it is for closed
systems.\cite{spanner_communication:_2010}  In addition, it has been
demonstrated that multiphoton effects can create large contribution
even in the linear absorption
regime,\cite{han_linear_2012,han_linear_2013,bruhl_experimental_2018,lavigne_ultrashort_2019}
which might lead to multiphoton phase control in seemingly one-photon
experiments.

In this paper, we show that the maximal time over which one-photon
phase control is stable is proportional to the length of the laser
pulse that produces the control.  That is, short laser pulses always
produce short-lived control, irrespective of the underlying system
dynamics.  This is a significant extension to previous work on the
impossibility of one-photon phase control in the steady
state.\cite{brumer_one_1989, spanner_communication:_2010,
mukamel_coherent-control_2013,am-shallem_scaling_2014}  Indeed, the
bound on control provided herein quantifies the ``steady state'' of
past results, in a fully general manner based on physical parameters.
Importantly, this bound gives support to the intuitive notion that
control in the steady state is not defined by an abstract
$t\rightarrow \infty$ limit; rather, \textit{the duration of the
control pulse is what bounds the duration of phase control
effects.}

Below, the physical conditions under which the bound on phase control is
established are carefully analyzed.  In doing so, a well-defined,
physically motivated and experimentally testable definition of
one-photon phase control is obtained.  Importantly, we propose, as did
others,\cite{han_linear_2012,han_linear_2013,bruhl_experimental_2018}
that weak multiphoton control is the source of the weak,
long-lived phase control previously measured in the linear regime in a
number of experimental and numerical investigations.\cite{
herek_quantum_2002,prokhorenko_coherent_2005, dietzek_mechanisms_2006,
prokhorenko_coherent_2006, katz_control_2010,
arango_communication:_2013}  Significantly, the novel, general bounds on
one-photon phase control provided herein strongly support a
multiphoton, rather than one-photon, phase control mechanism.

The results below are a direct consequence of the wave nature of the
classical electromagnetic field and of the time-translational
invariance usually assumed in nonlinear
spectroscopy.\cite{nuernberger_femtosecond_2009}  The derived bound on
the stability of phase control does not depend on the specific
dynamics of the controlled system.  Significantly, long-lived
one-photon phase control can not be stabilized by open quantum system
effects beyond the proposed bound.  In
particular, previously proposed environmentally-assisted phase control
mechanisms\cite{
  am-shallem_scaling_2014,
  lavigne_interfering_2017,bruhl_experimental_2018}
are insufficient in describing reported numerical and experimental
results.\cite{
herek_quantum_2002,prokhorenko_coherent_2005, dietzek_mechanisms_2006,
prokhorenko_coherent_2006, katz_control_2010,
arango_communication:_2013}

\section{Theory}

Phase control that is linear in the intensity of an exciting field is known to
be limited to time-dependent observables for time-independent initial
conditions in closed quantum mechanical systems.\cite{brumer_one_1989,
spanner_communication:_2010, mukamel_coherent-control_2013}  Below,
this result is extended to arbitrary quantum mechanical systems.  In
particular, oscillatory dynamics similar to those of
Eq.~(\ref{eq:example}) are shown to be a general feature even of complex
physical systems as a consequence of the wave nature of the electric
field and of time-translational invariance.

In the second section below, a significant extension to the theory of
one-photon phase control is obtained in the form of a strict bound on
the stability of control.  Specifically, phase control after absorption
of a laser pulse is shown to vanish over timescales much longer than
the duration of said pulse.  This bound quantifies previously
qualitative notions of ``long-lived'', ``stable'' or ``steady-state''
control.

\subsection{Time-translational invariance and phase control}

The interpretation of pulsed laser experiments fundamentally rely on
the important property of \textit{time-translational invariance} with
respect to the light.  That is, changing the time of arrival of the
exciting pulse only creates a translation of the observed dynamics ---
a measurement time $t$ due to a pulse at time $t'$ is a function of
the delay $t-t'$ only, as opposed to a more general two-dimensional
function of both $t$ and $t'$.  This property is a critical part of the
interpretation of spectroscopy experiments.  For example, a pump-probe
measurement with a delay $\tau$ between the pump and probe pulses is
obtained experimentally by the average of many repetitions, each with
the same delay.  At the core of such a protocol is the assumption that
the signal depends only on the delay $\tau$ between pump and probe
pulses, and not on their individual arrival times, i.e. that the
result is time-translational invariant.  In this section,
time-translational invariance is shown to result in oscillatory dynamics
when applied to one-photon phase control.

In general, the expectation value of a measurement of some property of a
quantum system arising from a one-photon excitation, e.g., the
population of a photoproduct, at some $t$ is given by a function of
two frequencies of the following form,
\begin{align}
  \braket{O(t)} \propto \iint_{-\infty}^\infty \mathrm d \omega_2  \mathrm d\omega_1 \ft\field(\omega_1)\ft\field(\omega_2) S_O(t, \omega_2, \omega_1).\label{eq:expectO}
\end{align}
where $S_O(t, \omega_2, \omega_1)$ is a two-frequency response
function for $O$.  The time dependence of $S_O(t,\omega_2,\omega_1)$
is oscillatory for a closed quantum mechanical system; this is derived
in Appendix~\ref{sec:pt} [Eq.~(\ref{eq:Opt})].  This result is not due
to the specific equation of motion of the controlled system; rather,
it is obtained as a direct consequence of the form of
Eq.~(\ref{eq:expectO}) and of time-translational invariance.  Consider
the electric field parameterized by a time shift $\tau$,
\begin{align}
  \field(t + \tau) = \field(t; \tau).
\end{align}
The Fourier transform of this expression is given by
\begin{align}
  \ft\field(\omega;\tau) = e^{i\omega \tau}\ft\field(\omega).\label{eq:shift-fourier}
\end{align}
Consider then Eq.~(\ref{eq:expectO}) but with the field and the
measurement of $O$ shifted forward in time by $\tau$,
\begin{align}
  \braket{O(t+\tau)} &\propto \int \mathrm d \omega_2 \int \mathrm d\omega_1 \ft\field(\omega_2;\tau)\ft\field(\omega_1;\tau) S_O(t+\tau, \omega_2, \omega_1)\\
  &= \int \mathrm d \omega_2 \int \mathrm d\omega_1 \ft\field(\omega_2)\ft\field(\omega_1) e^{i(\omega_1+\omega_2)\tau} S_O(t+\tau, \omega_2, \omega_1).\label{eq:osc-o}
\end{align}
If the underlying equations are time-translational invariant, a
measurement at $t+\tau$ due to a field shifted by $\tau$ must be
equivalent to a measurement at $t$ with no shift of the field.  If that
is the case for an arbitrary choice of the field, then
\begin{align}
 e^{i(\omega_1+\omega_2)\tau} S_O(t+\tau, \omega_2, \omega_1)= S_O(t, \omega_2, \omega_1),
\end{align}
which implies that $S_O(t,\omega_2, \omega_1) = S_O(0, \omega_2,
\omega_1) \exp(-i(\omega_1 + \omega_2) t)$.  That is, in accord with
Eq.~(\ref{eq:example}), the dynamics of $\braket{O(t)}$ is the
sum of oscillatory contributions with frequency
$\omega_1+\omega_2$ and amplitudes $\ft\field(\omega_1)\ft\field(\omega_2)$.

Measurements performed by a probe pulse, common in
experiments,\cite{
  prokhorenko_coherent_2005,prokhorenko_coherent_2006,
  prokhorenko_coherent_2011}
obey the exact same oscillatory dynamics.\cite{
mukamel_coherent-control_2013}  For example, the absorbance measured
with a probe pulse after excitation with a pump pulse (the
transient-absorption spectrum) is given by,
\begin{align}
    A_\text{pump-probe}(\omega_o, \tau_p) &= -\log\left(1 + \frac{I_{o,\text{probe}}(\omega_o) - I_{o,\text{pump-probe}}(\omega_o,\tau_p)}{|\ft\field_p(\omega_o)|^2}\right)
\end{align}
where $\tau_p$ is the delay between the pump and probe pulses,
$\ft\field_p(\omega_o)$ is the probe field at the measurement
frequency $\omega_o$, $I_{o,\text{probe}}(\omega_o)$ is the outgoing
field intensity of the probe after interaction with the system and
$I_{o,\text{pump-probe}}(\omega_o, \tau_p)$ is the same but subsequent to
an interaction with the pump.  This last term is the only term which
depends on $\tau_p$, and is given by the following nonlinear response,
\begin{align}
  I_\text{pump-probe}(\omega_o,\tau_p) &\propto \iiint \mathrm d \omega_3 \mathrm d \omega_2 \mathrm d \omega_1\; \ft\field^*_p(\omega_o; \tau_p)\ft\field_p(\omega_3;\tau_p) \ft\field(\omega_2;0)\ft\field(\omega_1;0)\nonumber \\
  &\quad\quad\times S_\mu(\tau_p, \omega_o, \omega_3, \omega_2, \omega_1)\label{eq:pump_probe}\\
                                       &=\iiint \mathrm d \omega_3 \mathrm d \omega_2 \mathrm d \omega_1\; e^{-i(\omega_o - \omega_3) \tau_p} \ft\field^*_p(\omega_o)\ft\field_p(\omega_3) \ft\field(\omega_2)\ft\field(\omega_1)\nonumber\\
  &\quad \quad \times  S_\mu(\tau_p,\omega_o, \omega_3, \omega_2, \omega_1)\nonumber
\end{align}
where $S_\mu(\tau_p,\omega_o,\omega_3,\omega_2,\omega_1)$ is a response
function, the specific form of which is derived in Appendix~\ref{sec:pt}.  In this case, the probe is centered at
$t=\tau_p$ and the pump is centered at $t=0$.  If time-translational invariance
holds, the result should be independent of an overall shift in time of
both the probe and the pump.  Consider then the same result but with
both pulse shifted backward by $\tau_p$ (i.e. with the probe and pump pulses centered at $t=0$ and $t=-\tau_p$ respectively),
\begin{align}
  I_\text{pump-probe}(\omega_o,\tau_p) &\propto \iiint \mathrm d \omega_3 \mathrm d \omega_2 \mathrm d \omega_1\; \ft\field^*_p(\omega_o; 0)\ft\field_p(\omega_3;0) \ft\field(\omega_2;-\tau_p)\ft\field(\omega_1;-\tau_p)\nonumber\\
  &\quad \times S_\mu(0, \omega_o, \omega_3, \omega_2, \omega_1)\nonumber\\
                                       &=\iiint \mathrm d \omega_3 \mathrm d \omega_2 \mathrm d \omega_1\; e^{-i(\omega_2 + \omega_1) \tau_p} \ft\field^*_p(\omega_o)\ft\field_p(\omega_3) \ft\field(\omega_2)\ft\field(\omega_1)\nonumber\\
  &\quad \times S_\mu(0, \omega_o, \omega_3, \omega_2, \omega_1).\label{eq:osc-pp}
\end{align}
where Eq.~(\ref{eq:shift-fourier}) has been used.  It thus follows
that, for a time-translational invariant system, a transient
absorption measurement will have the same oscillatory dynamics (with
oscillatory components $\omega_1+\omega_2$) as Eq.~(\ref{eq:osc-o})
above.  Thus, the formulas derived below equally apply to the
(experimentally important) case where phase control from a pump pulse
is measured by a probe pulse.

\subsection{Phase control dynamics after a pulsed excitation}

As shown in this section, the oscillatory dynamics [Eqs.~(\ref{eq:example}),
(\ref{eq:pump_probe}) and (\ref{eq:osc-pp})] of one-photon phase
control limit the possible duration of phase
control.  The dynamics of expectation values as obtained in
Eq.~(\ref{eq:expectO}) are the result of a two-dimensional frequency
integration over the field and the system's response.  For realistic
fields and systems, this dynamics can be extremely complex.  Yet, as
they are the result of a convolution with a time-limited pulse, phase
control is similarly time-limited.

\begin{figure}[h]
  \def\shift{-4.5}
\def\fieldsigma{2}
\begin{tikzpicture}[
  scale=1.0,
  interval/.style={densely dotted, thick},
  o1/.style={red},
  o2/.style={blue},
  ]

  \draw[interval] (4,\shift) -- (4,2.5) node[inner sep=0] (intone) {} node[left] {$t_0$};
  \draw[interval] (6.0,\shift) -- (6.0,2.5) node[inner sep=0] (inttwo) {} node[right] {$t_0 + T$};
  \draw[<->] (intone) -- (inttwo);

  \draw[o1] (-1,0) -- (0,0);
  \draw[o1,variable=\t,domain=0:6.5,smooth] plot[samples=100]
  ({\t}, {(1 + 0.4 * cos(360*\t)) * (1 - exp(-\t*\t / \fieldsigma))})
  node[right] {$\phi_1(\omega)$}; 

  \draw[o2] (-1,0) -- (0,0);
  \draw[o2,variable=\t,domain=0:6.5,smooth] plot[samples=100]
  ({\t}, {(1 + 0.4 * cos(360*\t + 140)) * (1 - exp(-\t *\t / \fieldsigma))})
  node[right] {$\phi_2(\omega)$}; 

  \draw[->] (-1,0) -- (8,0);
  \draw[->] (-1,0) -- (-1,2) node[above] {$\braket{O(t)}$};
  \draw[] (-2,1) node [] {\Large{A}};

  \draw[o1] (-1,\shift) -- (0,\shift);
  \draw[o1,variable=\t,domain=0:6.5,smooth] plot[samples=100]
  ({\t}, {( 1.5 + 0.2 * cos(360*\t * 1.2 + 150) ) * (1 - exp(-\t*\t / \fieldsigma))+\shift})
  node[right] {$\phi_1(\omega)$}; 

  \draw[o2] (-1,\shift) -- (0,\shift);
  \draw[o2,variable=\t,domain=0:6.5,smooth] plot[samples=100]
  ({\t}, {( 0.6 + 0.1 * cos(360*\t + 20) ) * (1 - exp(-\t *\t / \fieldsigma))+\shift})
  node[right] {$\phi_2(\omega)$}; 

  \draw[->] (-1,\shift) -- (8,\shift) node[below] {Time};
  \draw[->] (-1,\shift) -- (-1,2 + \shift) node[above] {$\braket{O(t)}$};
  \draw[] (-2,1 + \shift) node [] {\Large{B}};
  
\end{tikzpicture}
  \caption{Schematic representation of (A) short-lived, oscillatory
    control  and (B) long-lived control.  The expectation value
    of $\braket{O(t)}$ after excitation with a field
    $|\ft\field(\omega)|e^{i\phi_n(\omega)}$ is sketched.  Red and blue
    curves correspond to two fields differing only in the phase factor
    $\phi_n(\omega)$.  An
    averaging interval of size $T$ is shown by dotted line.}
  \label{fig:long-lived}
\end{figure}
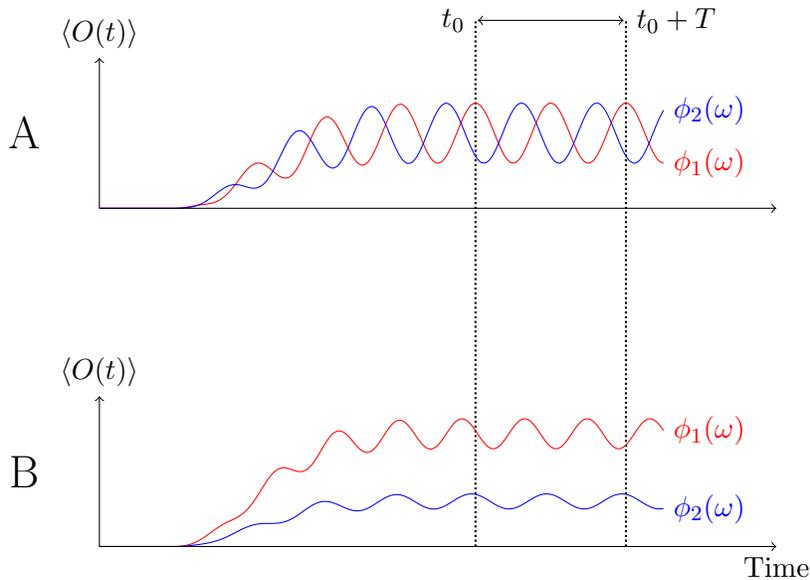

Frequencies $\omega_1$ and $\omega_2$ of the electric field contribute
to the dynamics of an observable a phase-controllable term which
oscillates with a frequency $\omega_1+\omega_2$.  Phase control over
long times can then be obtained from slowly oscillating terms (i.e.
where $\omega_1 \approx - \omega_2$).  Controlling such processes
should require precise manipulations of the phase of the field (i.e.
of the phase of $\ft\field(\omega_1)$ and $\ft\field^*(\omega_2)$).
However, as the uncertainty principle for $\omega$ and $t$ puts a
limit on the frequency resolution of the control field, phase control
of closely lying frequencies should be in some sense limited by the
duration of the field.  This is indeed correct, as demonstrated below.

First, the duration of phase control must be quantified.  One-photon phase control is only possible over
quantities exhibiting time-dependence.\cite{
  brumer_one_1989,spanner_communication:_2010,
  mukamel_coherent-control_2013}
The qualitative time-dependence obtained in closed quantum mechanical
system is demonstrated in Fig.~\ref{fig:long-lived}A, where the
expectation value $\braket{O(t)}$ of an observable $O$ is depicted for
two pulses (in red and blue) with differing phases.  For such systems,
the phase changes only the transient value of $\braket{O(t)}$.  In
contrast, phase control reported in Refs.~\onlinecite{
  herek_quantum_2002,prokhorenko_coherent_2005,
  prokhorenko_coherent_2006,katz_control_2010,
  arango_communication:_2013}
is remarkably stable,
with a time dependence qualitatively similar to that of
Fig.~\ref{fig:long-lived}B.  That is, phase control remains stable
over an extended time.

The time-average is used here to quantify the duration of control in a
way that can treat both regimes depicted in Fig.~\ref{fig:long-lived}.  By
averaging $\braket{O(t)}$ over an interval of time $T$, oscillations
much faster than $T$ are filtered out.\cite{lavigne_interfering_2017}
Phase control similar to that of Fig.~\ref{fig:long-lived}A can be
isolated from long-lived control similar to Fig.~\ref{fig:long-lived}B
by taking a sufficiently long yet finite time average; the value of
$T$ smoothly interpolates between the two regimes.  Below, the phase
control of an observable $O$ is said to be stable over $T$ if the
expectation value $\braket{O(t)}$ averaged over a time interval of
size $T$ after excitation depends on the phase of the exciting light.

The stability of control is shown in Appendix~\ref{sec:timecontrol} to
be bounded by the duration of a one-photon excitation.  The
time-averaged value of $\braket{O(t)}$ over a time interval from $t_0
-T/2$ to $t_0 +T/2$ is given by,
\begin{align}
  \overline O(t_0, T) = \frac{1}{T}\int_{t_0-T/2}^{t_0+T/2} \mathrm d t \braket{O(t)} = \overline O_\text{coh}(t_0, T) + \overline O_\text{inc}(t_0, T)\label{eq:coh-incoh-O}
\end{align}
where the coherent contribution $\overline O_\text{coh}(t_0, T)$
depends on the phase of the exciting field and the incoherent
contribution $\overline O_\text{inc}(t_0, T)$ does not.  These
contributions [Eq.~(\ref{eq:bound2}) in
Appendix~\ref{sec:timecontrol}] scale with the duration of the
exciting pulse $t_\field$ as,
\begin{align}
  \overline O_\text{coh}(t_0, T) \propto \frac{t_\field}{T} \text{ and } \overline O_\text{inc}(t_0, T) \propto \frac{T- t_\field}{T}.
\end{align}
for a time-limited field obeying $\field(t) = 0 $ for $|t| >
t_\field/2$ and where the averaging interval obeys $T>t_\field$.  A
derivation is provided in Appendix~\ref{sec:timecontrol}.  Thus, the
phase dependence of $\overline O(t_0, T)$ is directly related to the
duration of a pulsed excitation.  Similar bounds where previously
derived in the context of the short-time Fourier
transform\cite{slepian_prolate_1961,landau_prolate_1962} and as
extensions to the uncertainty principle for momentum and
position.\cite{j._b._m._uffink_uncertainty_1985}

In particular, the relations above impose a strict bound on stable
phase control in the approach to a steady-state, such as in the control of
photoisomerization.\cite{prokhorenko_coherent_2006}  The proportion of
control due to phase, i.e., the ratio of the coherent and incoherent
contributions in Eq.~(\ref{eq:coh-incoh-O}), has the following upper bound,
\begin{align}
  R_c = \max \frac{\overline O_{c}(t_0, T)}{\overline O(t_0, T)} = \frac{t_\field}{T} \frac{O_\text{max}}{O_\text{min}}.\label{eq:bound}
\end{align}
where $\braket{O(t)}$ is between values $O_\text{max}$ and
$O_\text{min}$ within the averaging interval.  In the approach to the
steady state (i.e. when $\braket{O(t)}$ has a functional dependence
similar to that depicted in Fig.~\ref{fig:long-lived}B), the ratio
$O_\text{max}/O_\text{min}$ is close to unity.  Then, phase control
vanishes when the pulse is significantly shorter than the interval
(when $t_\field \ll T$).  In particular, there is no one-photon phase
control of the population of quasi-steady states with control pulse
much shorter than the lifetime of said states, in direct contradiction
with a number of experiments and numerical simulations
on the control of isomerization and other molecular properties.\cite{prokhorenko_coherent_2005,prokhorenko_coherent_2006, dietzek_mechanisms_2006,katz_control_2010,arango_communication:_2013}

\section{Discussion}

As demonstrated, time-translational invariance sets bounds on the
amount of phase control that can be obtained from a one-photon
excitation.  This is in contrast with a number of experiments and
numerical studies that show steady- or near steady-state control in
the regime of linear absorption;\cite{
  herek_quantum_2002,prokhorenko_coherent_2005,
  prokhorenko_coherent_2006,katz_control_2010,
  arango_communication:_2013}
that is, $t_\field/T$ is very small but phase control is significant.
Below, previously advanced mechanisms for experimentally-obtained
long-lived one-photon phase control are reviewed and shown to
incompletely describe reported phase control.  To resolve this issue,
the set of assumptions required to obtain the bound on one-photon
phase control described above is carefully analyzed.  In particular,
the regime of one-photon phase control is rigorously defined.
Multiphoton or nonlinear phase control, occurring in the linear regime
of absorption, is proposed as the cause of the discrepancies between
this and other theoretical results \cite{
  brumer_one_1989, spanner_communication:_2010,
  mukamel_coherent-control_2013,am-shallem_scaling_2014}
on one side and experimental and numerical demonstrations of
long-lived one-photon phase control on the other side.\cite{
  herek_quantum_2002,prokhorenko_coherent_2005,
  prokhorenko_coherent_2006,katz_control_2010,
  arango_communication:_2013}

\subsection{Previously proposed mechanisms for one-photon phase control}

Previously proposed mechanisms in one-photon phase control are not
sufficient to explain the magnitude of reported long-lived phase
control results, as shown in this section.  Two such mechanisms have
previously been proposed: that transient control can be
long-lived\cite{spanner_communication:_2010,
  pachon_mechanisms_2013,arango_communication:_2013,
  lavigne_interfering_2017}
and that the environment can act to stabilize control in the manner of
a pump-dump experiment.\cite{
  prokhorenko_mechanism_2007,katz_control_2010,
  am-shallem_scaling_2014,lavigne_interfering_2017}
Neither mechanism allows violations of the bound described above.

First, experimentally reported long-lived phase control from
ultrashort pulses cannot be explained just as a consequence of
time-dependent but long-lived one-photon phase control, as was
previously done.\cite{spanner_communication:_2010,
pachon_mechanisms_2013,arango_communication:_2013,lavigne_interfering_2017}
For example, in the control of retinal
isomerization,\cite{prokhorenko_coherent_2006} phase effects from
pulses with duration $t_\field$ less than 2 ps were shown to be stable
over at least a 400 ps interval. Hence, the maximum \% change in the
isomer population from modifying only the phase of the field should be
$t_\field/T\approx 0.5\%$ .  This upper bound is significantly less
than the reported 2-4\% proportion of phase control.  Thus,
experimental control as reported cannot be entirely due to one-photon
phase effects, especially as the theoretical maximum described in this
paper does not account for experimental limitations (e.g. with
respect to pulse shaping) that further reduce the amount of available
control.\cite{lavigne_interfering_2017}  A similar analysis holds for
other
experiments.\cite{herek_quantum_2002,prokhorenko_coherent_2005,dietzek_mechanisms_2006,katz_control_2010,arango_communication:_2013}

Stabilization by an environment is the other major mechanism which has
been proposed to explain long-lived one-photon phase control.  That is,
in an open quantum system, relaxation of the one-photon induced
dynamics into a steady state could in some sense ``store'' one-photon
phase control, in a manner similar to a pump-dump experiment.\cite{
prokhorenko_mechanism_2007, katz_control_2010,
lavigne_interfering_2017,bruhl_experimental_2018} Physically however,
this mechanism does not respect time-translational invariance.  Here,
it is instructive to contrast qualitatively pump-dump control and the
proposed environmentally assisted pump-dump scheme.  A pump-dump
experiment occurs through two sequential light-matter
interactions,\cite{tannor_control_1985}
\begin{align}
  \ket{i} \xrightarrow{\ft\field_1(\omega)} \sum e^{-iE_i t}\ket{e_i} \xrightarrow{\ft\field_2(\omega)} \ket{f}.
\end{align}
The population of the final state can be shown in this case to be
phase-dependent.  For $\delta(t)$ pump and dump pulses, the expected
population of $\ket f$ will contain phase terms of the form
\begin{align}
  \sum_{i,j}\braket{e_j|\mu |f}\braket{f| \mu|e_i} e^{-i \omega_{ij} t_D},
\end{align}
where $t_D$ is the time between pump and dump pulses and $\mu$ is the
light-matter coupling operator.  In a pump-dump experiment, $t_D$ is fixed: it
is a control knob.  In an environmentally-assisted pump-dump scenario,
the ``dump'' can happen at any time, including before the pulse is on,
since the coupling with the environment does not change in
time.\footnote{
  If the system-environment coupling is time-dependent,
  control is possible but time-translational invariance is broken.  In
  fact, if the environment is taken to be a photon bath,
  environmentally-assisted pump-dump control simply becomes the usual
  pump-dump control scenario.}
No specific control can thus be gained, as the environment will dump
an oscillating coherence, but not at a well defined time.  Indeed,
the probability of dumping can only be controlled by the field 
through additional interactions with said field, i.e., through
interactions beyond those of one-photon absorption.  Control is then a
multiphoton effect.

Generally, the bound on phase control derived in this paper is not
specific to closed or Markovian dynamics.  Thus, environmentally
assisted phase control should obey the same bound; open system
dynamics do not allow for steady-state one-photon phase control.

\subsection{Conditions underlying the bound on one-photon phase control}

The bound on the stability of phase control was derived as the result
of a number of assumptions and approximations that are valid under
specific physical conditions.  A careful description of these
conditions follows, with tests of their regimes of validity.  These
conditions are separated into four sections: those arising from
semiclassical perturbation theory, those related to time-translational
invariance, those defining the one-photon regime and those related to
the neglect above of the spatial degrees of freedom of the electric
field.  A summary of some testable assumptions is given in Table
\ref{tab:tests}.

It should be noted that no claim is made about the possibility of
steady-state phase control or a signal masquerading as such.  That is,
phase control beyond the bound introduced in this article requires at
least one of the assumptions below to be false but no phase control is
guaranteed in any case.  The described tests are tools which can be
used to eliminate a range of possible mechanisms giving rise to a
phase control signal.

\begin{table}
\begin{ruledtabular}
  \begin{tabular}{lll}
    Assumption & Experimental variable & Effect on control \\
    \hline\hline
    Time-translational invariance & delay of $\field(t)$ & delay \\
    Identical repetitions & repetition rate & linear\\
    \hline
    Linear excitation & intensity of $\field(t)$  & linear\\
    Linear probe$^\dagger$ & intensity of $\field_p(t)$   & linear\\
    \hline
    No many-body interactions & concentration of sample   & linear\\
    No spatiotemporal coupling & measurement position & none \\
  \end{tabular}
\end{ruledtabular}
\begin{flushleft}
  {\footnotesize
  $\dagger$: For transient absorption and pump-probe measurements.}
\end{flushleft}
\caption{This table summarizes some assumptions taken in this paper,
as well as tests of their validity.  These tests are described in term
of the effect a change in a particular experimental variable should
have on the controlled signal if the corresponding assumption is
valid.  The exciting pulse is denoted $\field(t)$, while the probe
pulse in a pump-probe experiment is denoted $\field_p(t)$.  Details
are in the text. }\label{tab:tests}
\end{table}

\subsubsection*{Perturbative semiclassical description}

The transient dynamics of control are obtained based on the
description of a system and its environment evolving under Liouvillian
dynamics and interacting perturbatively with a classical field.  The
resulting theory has been consistently successful in the analysis of
nonlinear spectroscopy experiments,\cite{mukamel_principles_1995} and
coherent control experiments.\cite{shapiro_quantum_2012,
gordon_control_2015} As no additional restrictions are set on the
Liouvillian itself at this point, no assumptions of e.g. Markovian
dynamics, weak system-bath coupling, secular dynamics etc. is
required.\cite{lavigne_efficient_2019}  As such, the results of this article apply generally to open
and closed systems.  Finally, the dipolar approximation is taken, which
assumes that the molecule of interest is much smaller than the
wavelength of the exciting pulse, and that the important transitions
are dipole allowed.\cite{shapiro_principles_2003}

The semiclassical, perturbative approach forms the basis of the usual
description of coherent control and ultrafast
spectroscopy.\cite{lavigne_efficient_2019}  As such, no further
analysis of the above assumptions will be made.  An experimental
demonstration of one-photon phase control, in the regime of validity
of all the further assumptions and approximations described below,
would require a significant and interesting change in the way coherent
control and nonlinear spectroscopy are generally understood, but one
that is beyond the scope of this paper.

\subsubsection*{Time-translational invariance}

As described above, time-translational invariance and the Fourier
shift property guarantees that the phase control dynamics are given by
oscillatory components with amplitudes set by the electric field.
Time-translational invariance, the property that a delay of the field
leads only to a delay in the dynamics, is also an underlying
assumption of the interpretation of ultrafast spectroscopy
experiments.\cite{nuernberger_femtosecond_2009}  Broken
time-translational invariance might be caused by trivial effects, e.g.
laser drift, sample deterioration, etc. or possibly by non-trivial
physics such as time-dependent correlations between system and
bath.\cite{pachon_mechanisms_2013}  In any case, the result is a system
where a pump-probe experiment will depend on the time of the pump and
probe and not solely on the delay.

Closely related to time-translational invariance is the assumption
that a measurement of the expectation value to be controlled is the
result of an average over \textit{identical} repetitions.  Importantly
in ultrafast experiment, the sample is taken to either fully relax or
be fully renewed between pulses.  For systems with macroscopic
relaxation times, repeated interaction with pulsed lasers can create
non-trivial effects.\cite{paul_coherent_2017}  This assumption can be
tested by varying the repetition rate: if it holds, changing the
repetition rate will lead only to a change in the average power of the
beam at the sample, resulting in a linear change of the signal.

\subsubsection*{One-photon regime}

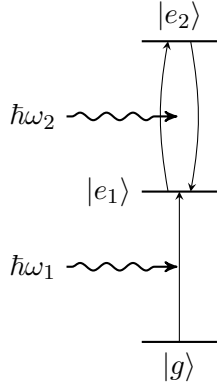
\begin{figure}[h]
  \begin{tikzpicture}[
  scale=0.5,
  decoration={snake, pre length=2pt, post length=5pt, amplitude=0.5mm},
  photon/.style={thick,->,decorate,>=stealth'},
  trans/.style={->,>=stealth},
  trans2/.style={double, <->,>=stealth},
  level/.style={thick},
  virtual/.style={thick, dashed},
  ]

  \draw[level] (-1,0) -- (1,0) node[below, midway] {$\ket{g}$};
  \draw[level] (-1,4) node[left] {$\ket{e_1}$}  -- (1,4) ;
  \draw[level] (-1,8) -- (1,8) node [above, midway] {$\ket{e_2}$};

  \draw[trans] (0,0) -- (0,4) node[midway, inner sep=0](w1) {} ;
  \draw[trans] (-0.3,4) to[bend left=10] (-0.3,8);
  \draw[trans] (0.3,8)  to[bend left=10] (0.3,4);
  \node[inner sep=0](w2) at (0, 6) {};

  \draw[photon] (w1)+(-3, 0) node[left] {$\hbar \omega_1$} -- (w1);
  \draw[photon] (w2)+(-3, 0) node[left] {$\hbar \omega_2$} -- (w2);
  
\end{tikzpicture}
  \caption{
    Example of a multiphoton process which can yield two-photon phase
    control in the linear regime of absorption.  The $\ket{g} \rightarrow
    \ket{e_1}$ one-photon absorption is modulated by a potentially
    phase-dependent $\ket{e_1} \leftrightarrow \ket{e_2}$ Raman
    transition, a two-photon correction to the one-photon
    absorption.\cite{faisal_theory_1987,konar_solvent_2014}}
  \label{fig:multi-from-one}
\end{figure}

The bound on phase control described in this paper is a consequence of
the specific forms of Eq. (\ref{eq:expectO}) for the expectation value
of an operator $O$.  In particular, the frequency integral for
$\braket{O(t)}$ includes a field dependence that is quadratic in the
amplitude $\ft\field(\omega)$ of the light.  The expectation value is then
linear in the intensity of $\ft\field(\omega)$, i.e., $\braket{O(t)}$ is
the result of a one-photon process.  Similarly, in a pump-probe or
transient-absorption experiment, the measured signal is also a
function of the intensity of the pump pulse $\ft\field(\omega)$
responsible for control.

Importantly, one-photon phase control is defined in this way as phase
control, linear in the intensity of the control field, of the value
of an observable.  This definition is significantly stricter than
qualitative notions of phase control ``in the linear regime'' or
``from a weak field.'' Multiphoton control of an observable $O$ is
possible in the linear regime of another observable $B$ or in the
linear regime of one-photon absorption.  For example, the two-photon
correction to the one-photon absorption, illustrated in
Fig.~\ref{fig:multi-from-one}, is phase-dependent.    Even if only 1\% of
the ground state $\ket{g}$ is excited, the two-photon correction could
modulate up to 1\% of the one-photon excited state $\ket{e_1}$.  Phase
control obtained from this two-photon correction in the linear regime
of excitation of $\ket{e_1}$ could easily be misconstrued for
steady-state one-photon phase control.

Previous work has shown that multiphoton effects of the kind shown in
Fig.~\ref{fig:multi-from-one} can be important even in the
low-intensity
regime.\cite{han_linear_2012,han_linear_2013,konar_solvent_2014,bruhl_experimental_2018}
Significantly, the linear regime may not be the same for all pulse
shapes and for all observables.  For example, consider the control of
isomer populations in rhodopsin.  It may be convenient to establish
linearity with respect to absorption, but that does not guarantee
linearity with respect to the control targets, the isomer populations,
as shown above.  Hence, it is critical to demonstrate linearity for the
\textit{controlled observable} with respect to the intensity of the
\textit{control pulse.}

An additional complication is present in pump-probe and transient
absorption experiments.  In this case, the signal is taken to be the
result of a first-order interaction with the probe pulse detected by
heterodyning, after a one-photon interaction with the pump
pulse.\cite{mukamel_principles_1995}  In particular, higher-order
effects due to the probe are neglected, as are any signals generated
by the pump alone (e.g. pump-induced Raman scattering in the probe
direction).  The signal of a one-photon phase control experiment should
be linear with the intensity of the probe pulse, in addition to being
linear with the intensity of the control pulse.  It should be noted
that this is a stronger condition than the signal being quadratic in
the intensity of the combined pump and probe field, which includes,
for example, two-photon processes due to the pump field with no
contribution from the probe field.  Linearity of the pump-probe signal
must be separately tested for both pump and probe fields.

\subsubsection*{Spatial dependence of the field}

Finally, interference between spatial components of the field and of
the sample could produce phase-dependent effects that can be mistaken for
one-photon phase control.  Firstly, it is assumed here that a
change of the phase of the field does not modify the beam profile or
the intensity of the field.  The converse, spatiotemporal coupling, is
a well-known classical interference effect.  Phase control of
spatiotemporal coupling, which has previously been
demonstrated,\cite{katz_focusing_2011,mounaix_spatiotemporal_2016}
could produce signals analogous to one-photon phase control.  Such
effects can be mitigated by careful experimental
design.\cite{liebel_lack_2017}  In particular, in the absence of
spatiotemporal coupling, the field is uniform over the sample.  Then,
shifting measurement volume or the excitation volume should have no
effect on the signal.\cite{brinks_beating_2011}

Secondly, phase-dependent collective interactions of
multiple molecules with radiation are also neglected.  Experimentally,
this is the case if the sample is sufficiently transparent and
many-body interactions can be neglected, which can be tested by
verifying that the properties of individual absorbers within the
sample are independent of the presence of other absorbers.  If that is
the case, a change in the concentration of the sample will be
reflected simply by a linear change in the signal, as only the number
of absorbers is changed. 

\section{Conclusion}
 
In this paper, we demonstrated that one-photon phase control is
always transient, in accordance with previous theoretical
results.\cite{brumer_one_1989,spanner_communication:_2010,mukamel_coherent-control_2013}
Phase control stable over an interval much longer than the duration of
exciting light has been reported in several experimental and numerical
investigations. \cite{herek_quantum_2002,prokhorenko_coherent_2005,dietzek_mechanisms_2006,katz_control_2010,arango_communication:_2013}
These results are in contradiction with the bound on one-photon phase
control given in this paper; this inconsistency is not resolved by
previously proposed
mechanisms.\cite{prokhorenko_coherent_2006,katz_control_2010,arango_communication:_2013,lavigne_interfering_2017,bruhl_experimental_2018}

The relationship between the dynamics of one-photon phase control and
the duration of the exciting pulse is a significant and novel
extension of past work, where time-dependence was found as a condition
for one-photon phase control, but was not
quantified.\cite{spanner_communication:_2010,mukamel_coherent-control_2013}
The bound in Eq.~(\ref{eq:bound}) provides a quantitative limit for
the previously proposed mechanism describing stable phase control
qualitatively as the result of transient but long-lived effects.

Phase control violating this bound cannot be explained by open quantum
system dynamics.  The results of this article are fully general and no
assumptions of e.g. Markovianity or weak system-bath coupling were
made.  Hence, the proposed mechanism, by which a non-Markovian
environment acts in a manner similar to a dump pulse in a pump-dump
control scenario, cannot account for reported steady-state one-photon
phase control.  This work thus extends theoretical results showing the
impossibility of such control in scattering and Markovian
dynamics\cite{brumer_one_1989,am-shallem_scaling_2014} to all
time-translationally invariant systems.

Conditions for this result were carefully examined, and tests of their
validity were proposed.  Long-lived control with an ultrashort pulse
was shown to require at least one such condition to be broken.  The
proposed tests can be used as a tool to characterize the mechanism of
phase control.

We hypothesize that multiphoton effects are responsible for reported
long-lived phase control results, as was previously
proposed.\cite{han_linear_2012,han_linear_2013,liebel_lack_2017,
bruhl_experimental_2018}  Phase-dependent multiphoton effects can
appear in linear regime experiments and numerical simulations.  For
example, phase-dependent higher order corrections (e.g., as shown in
Fig.~\ref{fig:multi-from-one} and experimentally measured in
Ref.~\onlinecite{konar_solvent_2014}) can modulate the linear
absorption in a way which may be difficult to distinguish from
one-photon phase control, and may cause significant phase effects even
at low intensity.  For example, if only 2\% of the sample is excited,
it may be convenient to assume that multiphoton effects can be
neglected. However, the excited portion of the sample contributes the
whole of the signal; 2\% of the signal can well be the result of
two-photon processes.  The resulting two-photon phase control can be
easily be misconstrued for small but measurable one-photon phase
control.  Indeed, the low field intensities used in some experiments
does not guarantee that phase control is due to one-photon term.  The
small magnitude of phase control reported in e.g.,
Refs.~\onlinecite{prokhorenko_coherent_2005,prokhorenko_coherent_2006}
is consistent with the expected small magnitude of two-photon
processes in the linear regime of absorption.

Multiphoton effects are likely the cause of stable phase
control in simulations not performed in the framework of perturbation
theory.\cite{katz_control_2010, arango_communication:_2013}  For
example, in Ref.~\onlinecite{arango_communication:_2013}, a simulation
using MCTDH shows phase control in retinal isomerization using a
closed quantum model consisting of a large number of vibrational
modes.  However, the intensity is sufficiently high as to include
significant multiphoton contributions.  The nonlinearity in this case
is not the same for all excitation pulses, and is in fact phase
dependent.\footnote{
  This is clearly demonstrated by the fact that at the smallest
  simulated intensity, no difference between positively and negatively
  chirped pulses is detected.  In the linear regime, a change in
  intensity should create a proportional change in the signal of both
  positively and negatively chirped pulses: as such, the ratio of the
  signal from the positively and negatively chirped pulses should remain
  constant.}
In contrast, explicit perturbation theory computations where only
one-photon processes are allowed do not show long-lived control, in
agreement with the results of this
paper.\cite{christopher_overlapping_2005,christopher_quantum_2006,
pachon_mechanisms_2013,am-shallem_scaling_2014,lavigne_interfering_2017}
(An efficient method to compute the perturbative series obtained in
ultrafast experiments is given in
Ref.~\onlinecite{lavigne_efficient_2019}. )

Multiphoton effects as the source of phase control would resolve
a number of issues raised by multiple authors on the feasibility and
mechanism of one-photon phase control,\cite{
  joffre_comment_2007,prokhorenko_response_2007,
  nuernberger_femtosecond_2009,han_linear_2012,
  han_linear_2013, liebel_lack_2017,bruhl_experimental_2018}
and is fully in accord with the analytical
results of this article and others on the transient dynamics of
control.\cite{
  brumer_one_1989, spanner_communication:_2010,
  mukamel_coherent-control_2013,am-shallem_scaling_2014}
In addition, multiphoton phase control is well-understood from a
physical point of view.\cite{shapiro_quantum_2012}  This
mechanism provides a solid starting point for the identification and
exploitation of new control schemes.  In contrast, mechanisms that
allow for long-lived one-photon phase control are qualitative at best
(e.g. with control identified solely as the result of ``non-Markovian
effects''). 

Significantly, long-lived one-photon phase control is incongruent
with the theory of nonlinear spectroscopy, which shares the same
assumptions used in this
paper.\cite{yan_optical_1993,mukamel_principles_1995,mukamel_coherent-control_2013}
The described tests above can eliminate a range of alternative
explanations for stable, linear phase-dependent control results.  The
remaining effects, for which no tests where given, include the
presence of quantum optical effects, the breakdown of the light-matter
perturbative expansion and dynamics not accounted in the Liouville
equation.  Control experiments in contradiction with the results of
presented here are performed in much the same way as other nonlinear,
ultrafast spectroscopy
experiments.\cite{herek_quantum_2002,prokhorenko_coherent_2005,prokhorenko_coherent_2006}
Thus, if these somewhat esoteric effects have a significant impact in
coherent control experiments, their inclusion may also be required in
the interpretation of other condensed phase ultrafast experiments.

\textbf{Acknowledgements:} This work was supported by the
U.S. Air Force Office of Scientific Research under Contract No.
FA9550-17-1-0310, and by the
Natural Sciences and Engineering Research Council of Canada.

\bibliography{onephoton}

\appendix
\section{Light-matter interaction for a general quantum mechanical system}
\label{sec:pt}

In this section, the interaction of a quantum mechanical system with
radiation is derived in the usual way.\cite{mukamel_principles_1995}
The system without radiation, represented by a density matrix $\rho(t,
\vec r)$, evolves under the action of a generally time-dependent
Liouville operator,
\begin{align}
  \frac{\mathrm d }{\mathrm d t}\rho(t) = \super L_0 (t) \rho(t).
\end{align}
Here, the Liouvillian consists of an absorber, its environment and the
coupling between them.  That is, a separation is made between the field
and the system but the system itself is not partitioned into system
and environment.  In the dipole approximation, the semiclassical
light-matter interaction Liouvillian is given by
\begin{align}
    \frac{\mathrm d }{\mathrm d t}\rho(t) = \super L_0 (t) \rho(t) + \vec \field(t, \vec r) \cdot \super V \rho(t).
\end{align}
where $i\hbar\super V \rho = \commute{\vec \mu}{\rho}$, $\vec \mu$ is
the dipole operator and $\vec \field(t,\vec r)$ is the electric field
at the center of charge $\vec r$ the molecule.  Below, the macroscopic spatial dependence
of the electric field is dropped, which corresponds
physically to the case of an homogeneous isotropic and near
transparent sample.\cite{shapiro_quantum_2012}

The system is taken to evolve under equations that are
invariant under time translations of the electric field.
Hence, a shift of $\tau$ of the field must lead to the same dynamics
but shifted by $\tau$,
\begin{align}
  \super L_0 (t) \rho(t) + \field(t) \super V \rho(t) &= \super L_0 (t) \rho(t+\tau) + \field(t+\tau) \super V \rho(t+\tau)\nonumber\\
                                                    &= \super L_0 (t-\tau) \rho(t) + \field(t) \super V \rho(t).\nonumber
\end{align}
Therefore the following must hold,
\begin{align}
  \super L_0(t) = \super L_0(t-\tau).
\end{align}
Time-translational invariance broken only by the field therefore
implies that the field-free Liouvillian is time-independent.  (This is
in addition to the initial state being a steady state.)  An
interaction picture is defined by,
\begin{align}
  \super U_0(t) &= \exp\left[\super L_0 t\right]\\
  \rho_I(t) &= \super U_0(-t) \rho(t)\\
  \super V_I(t) &= \super U_0(-t) \super V \super U_0(t).
\end{align}
Using these quantities, the equation of motion
in the interaction picture becomes 
\begin{align}
  \frac{\mathrm d} {\mathrm d t} \rho_I(t) = \field(t)\super V_I(t)\rho_I(t)
\end{align}
Formal integration yields the Dyson series,
\begin{align}
  \rho_I(t) &= \rho(t_0) + \int_{t_0}^t \mathrm d t_1 \field(t_1) \super V_I(t_1) \rho_I(t_0) \\
  &\qquad + \int_{t_0}^t \mathrm d t_2 \int_{t_0}^{t_2} \mathrm d t_1 \field(t_2) \field(t_1) \super V_I(t_2) \super V_I(t_1) \rho_I(t_0) + O(\field(t)^3).\nonumber
\end{align}

Processes that are linear in the field intensity result from the above
perturbative expansion truncated at second order,
\begin{align}
  \rho_I(t) &= \rho_I(t_0) + \int_{t_0}^t \mathrm d t_1 \field(t_1) \super V_I(t_1) \rho_I(t_0) \\
  &\qquad + \int_{t_0}^t \mathrm d t_2 \int_{t_0}^{t_2} \mathrm d t_1 \field(t_2) \field(t_1) \super V_I(t_2) \super V_I(t_1) \rho_I(t_0).\nonumber
\end{align}
Transforming back to the Schrodinger picture gives,
\begin{align}
  \rho(t)& = \super U_0(t-t_0) \rho(t_0) + \int_{t_0}^t \mathrm d t_1 \field(t_1) \super U_0 (t-t_1)\super V \super U_0(t_1 -t_0) \rho(t_0)\label{eq:rt} \\
  &\qquad + \int_{t_0}^t \mathrm d t_2 \int_{t_0}^{t_2} \mathrm d t_1 \field(t_2) \field(t_1) \super U_0(t-t_2)\super V \super U_0(t_2-t_1) \super V \super U_0(t_1-t_0) \rho(t_0).\nonumber
\end{align}
The time $t_0$ is taken to be before the field is on.
Time-translational invariance implies that $\rho(t_0)$ is a
time-independent steady state, that is
\begin{align}
  \super U_0(t)\rho(t_0) = \rho(t_0) = \rho_0.
\end{align}
The causal Green's function is defined as,
\begin{align}
  \super G_0(t) = \Theta(t) \super U_0(t) = \Theta(t) \exp(\super L_0 t).
\end{align}
where $\Theta(t)$ is the Heaviside step function.  Eq.~(\ref{eq:rt}) can then
be expressed as
\begin{align}
  \rho(t)& = \rho_0 + \int_{-\infty}^\infty \mathrm d t_1 \field(t_1) \super G_0 (t-t_1)\super V \rho_0 \label{eq:pt-td}\\
  &\qquad + \int_{-\infty}^\infty \mathrm d t_2 \int_{-\infty}^\infty \mathrm d t_1 \field(t_2) \field(t_1) \super G_0(t-t_2)\super V \super G_0(t_2-t_1) \super V \rho_0\nonumber
\end{align}
where $t_0$ has been taken to $-\infty$.  The expansion to higher order
is of a similar form.  

\subsection*{Frequency domain expressions}

As the phase is easily expressible only in the frequency domain,
coherent control is fundamentally spectral in nature.  The inverse
Fourier transform of the field is given by
\begin{align}
  \field(t) = \frac{1}{2\pi} \int_{-\infty}^\infty \mathrm d \omega e^{+i\omega t}  \ft\field(\omega)
\end{align}
Applying this expression to Eq.~(\ref{eq:pt-td}) yields
\begin{align}
  \rho(t) = \rho_0 + &\left(\frac{1}{2\pi}\right) \int_{-\infty}^\infty \mathrm d \omega_1 \ft\field(\omega_1) \int_{-\infty}^\infty \mathrm d t_1 e^{i \omega_1 t_1}\super G_0 (t-t_1)\super V \rho_0 \\
  + &\left(\frac{1}{2\pi}\right)^2\iint_{-\infty}^\infty \mathrm d \omega_1 \mathrm d \omega_2 \ft\field(\omega_2) \ft\field(\omega_1) \iint_{-\infty}^\infty \mathrm d t_2 \mathrm d t_1  e^{i\omega_1t_1 + \omega_2t_2} \super G_0(t-t_2)\super V \super G_0(t_2-t_1) \super V \rho_0\nonumber
\end{align}
The Fourier transforms over $\super G_0(t)$ are given by,
\begin{align}
  \int_{-\infty}^\infty \mathrm d\tau e^{i \omega (t-\tau)}\super G_0 (\tau) = \lim_{\epsilon \rightarrow 0^+} e^{i(\omega - i\epsilon) t}\ft{\super G}_0(\omega).
\end{align}
with the Green's function $\ft{\super G}_0(\omega)
= \left[i\omega - ( \super L_0 + \epsilon )\right]^{-1} $.  (An alternative derivation based on the Laplace transform is given in Ref.~\onlinecite{lavigne_efficient_2019}.)  Then, the
perturbative expansion becomes,
\begin{align}
  \rho(t) = \rho_0 &+ \left(\frac{1}{2\pi}\right)\int_{-\infty}^\infty \mathrm d \omega_1 e^{i\omega_1 t + \epsilon t}\ft\field(\omega_1) \ft{\super G}_0 (\omega_1)\super V \rho_0 \\
  &+ \left(\frac{1}{2\pi}\right)^2\iint_{-\infty}^\infty \mathrm d \omega_2  \mathrm d \omega_1 e^{i(\omega_2 + \omega_1) t + \epsilon t}\ft\field(\omega_2) \ft\field(\omega_1) \ft{\super G}_0(\omega_2+ \omega_1)\super V \ft{\super G}_0(\omega_1) \super V \rho_0\nonumber.
\end{align}
The electric field $\field(t)$ is zero for $t < C$ for some $C$.  Then,
the above integrals converge at any time $t$ in the $\epsilon
\rightarrow 0^+$ limit as a consequence of the Paley-Wiener
theorem.\cite{rudin_functional_1991}  To simplify the notation, factors
of $\epsilon$ are made implicit below.

The third order expansion, which will be of use for pump-probe
measurements, can be derived in much the same way,
\begin{align}
    \rho(t) = \rho_0  +  &\left(\frac{1}{2\pi}\right)\int_{-\infty}^\infty \mathrm d \omega_1 e^{i\omega_1 t }\ft\field(\omega_1) \ft{\super G}_0 (\omega_1)\super V \rho_0 \label{eq:thirdorder}\\
            + &\left(\frac{1}{2\pi}\right)^2\iint_{-\infty}^\infty \mathrm d \omega_2 \mathrm d \omega_1 e^{i(\omega_2 + \omega_1) t}\ft\field(\omega_2) \ft\field(\omega_1) \ft{\super G}_0(\omega_2+ \omega_1)\super V \ft{\super G}_0(\omega_1) \super V \rho_0\nonumber\\
            + &\left(\frac{1}{2\pi}\right)^3\iiint_{-\infty}^\infty \mathrm d \omega_3 \mathrm d \omega_2  \mathrm d \omega_1 e^{i(\omega_3 + \omega_2 + \omega_1) t}\ft\field(\omega_3) \ft\field(\omega_2) \ft\field(\omega_1)\nonumber\\
  &\qquad\times\ft{\super G}_0(\omega_3 + \omega_2 + \omega_1)\super V \ft{\super G}_0(\omega_2+ \omega_1)\super V \ft{\super G}_0(\omega_1) \super V \rho_0\nonumber.             
\end{align}
The expectation value of an operator can be written using the above in
the form of frequency correlation functions,
\begin{align}
  O(t) = O^{(0)} + &\left(\frac{1}{2\pi}\right)\int_{-\infty}^\infty \mathrm d \omega_1 e^{i\omega_1 t }\ft\field(\omega_1) S_O^{(1)}(\omega_1) \label{eq:Opt}\\
            + &\left(\frac{1}{2\pi}\right)^2\iint_{-\infty}^\infty \mathrm d \omega_2 \mathrm d \omega_1 e^{i(\omega_2 + \omega_1) t}\ft\field(\omega_2) \ft\field(\omega_1) S_O^{(2)}(\omega_2,\omega_1)\nonumber\\
            + &\left(\frac{1}{2\pi}\right)^3\iiint_{-\infty}^\infty \mathrm d \omega_3 \mathrm d \omega_2  \mathrm d \omega_1 e^{i(\omega_3 + \omega_2 + \omega_1) t}\ft\field(\omega_3) \ft\field(\omega_2) \ft\field(\omega_1)S_O^{(3)}(\omega_3,\omega_2,\omega_1) \nonumber
\end{align}
where,
\begin{align}
  S_O^{(0)} &= \Tr \left[O \rho_0 \right]\\
  S_O^{(1)}(\omega_1) &= \Tr \left[O\ft{\super G}_0 (\omega_1)\super V \rho_0\right]\\
  S_O^{(2)}(\omega_2,\omega_1) &= \Tr \left[O\ft{\super G}_0(\omega_2+ \omega_1)\super V \ft{\super G}_0(\omega_1) \super V \rho_0\right]\\
  S_O^{(3)}(\omega_3,\omega_2,\omega_1) &= \Tr \left[ O\ft{\super G}_0(\omega_3 + \omega_2 + \omega_1)\super V \ft{\super G}_0(\omega_2+ \omega_1)\super V \ft{\super G}_0(\omega_1) \super V \rho_0 \right].
\end{align}

\subsection*{Pump-probe and transient absorption spectroscopy}

Transient absorption and pump-probe spectroscopy measure the action of
a pump pulse, i.e. the pulse used to control the system, on the
absorption of a weak probe pulse.  The intensity of the probe pulse
after interaction with the system at a measured output frequency $\omega_o$ is
given by,\cite{mukamel_coherent-control_2013}
\begin{align}
  I_o(\tau_p, \omega_o) &\propto |\ft\field_p(\omega_o;\tau_p) + \ft\field_s(\omega_o)|^2\\
                        &= |\ft\field_p(\omega_o;\tau_p)|^2 + |\ft\field_s(\omega_o)|^2 + 2\Real\left[\ft\field^*_p(\omega_o;\tau_p)\ft\field_s(\omega_o)\right]
\end{align}
where $\ft\field_p(\omega_o;\tau_p)$ is the probe field with delay
$\tau_p$ and $\ft\field_s(\omega_o)$ is the signal field outgoing from
the system.  The signal field is proportional to the polarization,
phase shifted by $\pi/2$.  For a weakly polarized system, consistent
with a near transparent sample, only the heterodyne signal is
significant.  The change in intensity of the probe pulse after
interaction with the sample is then given by,
\begin{align}
  \Delta I(\tau_p, \omega_o) = I_o(\tau_p, \omega_o)-I_p(\omega_o) &\propto -\Imag\left[\ft\field^*_p(\omega_o;\tau_p) \mu(\omega_o)\right]
\end{align}
where $I_p(\omega_o) = |\ft\field_p(\omega_o;\tau_p)|^2$ and
$\mu(\omega_o)$ is the Fourier transform of the dipole expectation
value $\braket{\mu(t)}$.

The change in intensity of the probe given prior one-photon absorption
of a pump pulse is given by the following four-wave mixing
contribution,
\begin{align}
  \Delta I_\text{pump-probe}(\tau_p, \omega_o) &\propto -\Imag \bigg[\ft\field^*_p(\omega_o;\tau_p) \int_{-\infty}^{\infty} \mathrm d t e^{-i\omega_o t} \iiint_{-\infty}^\infty \frac{\mathrm d \omega_3}{2\pi} \frac{\mathrm d \omega_2}{2\pi} \frac{\mathrm d \omega_1}{2\pi} e^{i (\omega_3 + \omega_2 + \omega_1) t} \\
  &\qquad\qquad\times \ft\field_p(\omega_3;\tau_p)\ft\field(\omega_2) \ft\field(\omega_1) S_\mu^{(3)}(\omega_3 ,\omega_2,\omega_1)\bigg]\nonumber\\
  &\propto -\Imag \bigg[ \iint_{-\infty}^\infty \frac{\mathrm d \omega_2}{2\pi} \frac{\mathrm d \omega_1}{2\pi} \ft\field^*_p(\omega_o;\tau_p) \ft\field_p(\omega_o - \omega_2 - \omega_1;\tau_p)\\
  &\qquad\qquad\times \ft\field(\omega_2) \ft\field(\omega_1) S_\mu^{(3)}(\omega_o-\omega_2 -\omega_1 ,\omega_2,\omega_1)\bigg].\nonumber
\end{align}
The delay $\tau_p$ can be converted to an oscillation frequency using
the Fourier shift theorem [i.e. as a consequence of
Eq.~(\ref{eq:shift-fourier})] to obtain,
\begin{align}
  \Delta I_\text{pump-probe}(\tau_p, \omega_o) &\propto -\Imag \bigg[\iint_{-\infty}^\infty \frac{\mathrm d \omega_2}{2\pi} \frac{\mathrm d \omega_1}{2\pi} e^{-i(\omega_2 +\omega_1 ) \tau_p } \ft\field^*_p(\omega_o) \ft\field_p(\omega_o - \omega_2 - \omega_1)\\
  &\qquad\qquad\times \ft\field(\omega_2) \ft\field(\omega_1) S_\mu^{(3)}(\omega_o-\omega_2 -\omega_1 ,\omega_2,\omega_1)\bigg].\nonumber
\end{align}
This form has equivalent dynamics as Eq.~(\ref{eq:Opt}) above but with
$\tau_p$ taking the role of $t$.

\section{Dynamics of one-photon phase control} \label{sec:timecontrol}
In this section, dynamical bounds on one-photon phase control from a
pulse of light are derived.  This bound is broadly similar to that of
Slepian and Pollak for the resolution of the short-time Fourier
transform\cite{slepian_prolate_1961} as well as the result of Uffink
and Hilgevoord\cite{j._b._m._uffink_uncertainty_1985} on the
resolution of the double slit experiment.  The field is taken to be
time-limited, with
\begin{align}
  \field(t') = 0  \text{ for } |t'| > t_\field.
\end{align}
The expectation value of an observable $O$ at a time $t>t_\field$ is
given by the following general formula,
\begin{align}
  \braket{O(t)} &= \int_{-\infty}^\infty \mathrm d t_2 \int_{-\infty}^{\infty} \mathrm d t_1 \field(t_1) \field(t_2) S_O(t-t_2, t_2-t_1)
\end{align}
where $S_O(t-t_2, t_2-t_1)$ is a function which gives the response of
$\braket{O(t)}$.  Specifically, $S_O(t-t_2, t_2-t_1)$ is given by,
\begin{align}
  S_O(t-t_2, t_2-t_1) = \Tr \left[O\super G_0(t-t_2) \super V \super G_0(t_2 - t_1) \super V \super \rho_0\right]
\end{align}
which can be obtained by taking the expectation value of $O$ using the
second term of Eq.~(\ref{eq:pt-td}) above.  

The response function can be expressed as sum of a $t$-independent
term, $\overline S_O(t_2-t_1)$ and a $t$-dependent term $\delta
S_O(t-t_2, t_2-t_1)$.  The following is then obtained,
\begin{align}
  \braket{O(t)} &= \int_{-\infty}^\infty \mathrm d t_2 \int_{-\infty}^{\infty} \mathrm d t_1 \field(t_1) \field(t_2) \left(\overline S_O(t_2-t_1) +  \delta S_O(t-t_2, t_2-t_1)\right).
\end{align}
Expressing the fields in the frequency domain yields
\begin{align}
  \braket{O(t)} &= \iint_{-\infty}^\infty \frac{\mathrm d \omega_2}{2\pi} \frac{\mathrm d \omega_1}{2\pi} \ft\field(\omega_1) \ft\field^*(\omega_2) \\
  &\quad \times \iint_{-\infty}^{\infty} \mathrm d t_2  \mathrm d t_1 e^{i\omega_1 t_1 - i \omega_2 t_2} \left[\overline S_O(t_2-t_1) + \delta S_O(t-t_2, t_2-t_1)\right].\nonumber
\end{align}
The substitution $\tau = t_2-t_1$ is performed to obtain,
\begin{align}
  \nonumber &\iint_{-\infty}^{\infty} \mathrm d t_2  \mathrm d \tau e^{i(\omega_1-\omega_2)t_2 - i\omega_1\tau} \left[\overline S_O(\tau) + \delta S_O(t-t_2, \tau)\right]\\
  \nonumber &= \overline S_O(\omega_1)2\pi\delta(\omega_1-\omega_2) + \int_{-\infty}^{\infty} \mathrm d t_2 e^{i(\omega_1-\omega_2)t_2} \delta S_O(t-t_2, \omega_1)\\
  &= \overline S_O(\omega_1)2\pi\delta(\omega_1-\omega_2) + e^{i(\omega_1-\omega_2)t}\int_{-\infty}^{\infty} \mathrm d \tau e^{-i(\omega_1-\omega_2)\tau} \delta S_O(\tau, \omega_1)  \label{eq:Ofield}
\end{align}
The expectation value can thus be written as
\begin{align}
  \braket{O(t)} &= \int_{-\infty}^\infty \frac{\mathrm d \omega_1}{2\pi} |\ft\field(\omega_1)|^2 \overline S_O(\omega_1) \\
  \nonumber&\quad + \iint_{-\infty}^\infty \frac{\mathrm d \omega_1}{2\pi}  \frac{\mathrm d \omega_2}{2\pi} e^{i(\omega_1-\omega_2)t}\ft\field(\omega_1)\ft\field^*(\omega_2) \delta S_O(\omega_1-\omega_2, \omega_1).
\end{align}
The first term gives the steady-state contribution and is phase
insensitive.  The second term is phase sensitive; however, it is also
time-dependent, as expected.  The above derivation reproduces the
results of
Refs.~\onlinecite{brumer_one_1989,spanner_communication:_2010,mukamel_coherent-control_2013},
but does not quantify the relationship between time-dependence and
phase control.

Here, an upper bound on the stability of phase control (defined using
the time average) is obtained based on the duration of the exciting
field.  First, equation (\ref{eq:Ofield}) is used without the
separation into steady and non-steady contributions,
\begin{align}
  \braket{O(t)} &= \iint_{-\infty}^\infty \frac{\mathrm d \omega_1}{2\pi} \frac{\mathrm d \Omega}{2\pi} \ft\field(\omega_1)\ft\field^*(\omega_1 + \Omega) \int_{-\infty}^{\infty} \mathrm d \tau e^{-i\Omega(\tau - t)} S_O(\tau, \omega_1).
\end{align}
where the change of variable $\omega_2 = \omega_1 + \Omega$ has been
effected. $\braket{O(t)}$ can approach a steady state if such state
exists, but it can also be oscillatory (as is the case for a closed
system) at all times.  The stability of control and the approach to
the steady state are described by taking a time
average,
\begin{align}
  \overline O(t_0,2T) = \frac{1}{2T}\int_{t_0-T}^{t_0 + T}\mathrm d t \braket{O(t)}.
\end{align}
Thus, $\overline O(t_0,2T)$ describes the value of $O$ averaged over
the window from $t_0-T$ to $t_0 + T$, where $t_0 - T > t_\field$.  It
is given by
\begin{align}
  \nonumber\overline O(t_0,2T) &= \frac{1}{2T}\int_{t_0-T}^{t_0 + T}\mathrm d t \iint_{-\infty}^\infty \frac{\mathrm d \omega_1}{2\pi} \frac{\mathrm d \Omega}{2\pi} \ft\field(\omega_1)\ft\field^*(\omega_1 + \Omega) \int_{-\infty}^{\infty} \mathrm d \tau e^{-i\Omega(\tau - t)} S_O(\tau, \omega_1).
\end{align}
The $\Omega$ integral is performed first, inverting one of the Fourier
transforms,
\begin{align}
  \nonumber\overline O(t_0,2T) &= \frac{1}{2T} \int_{-\infty}^\infty\frac{\mathrm d \omega_1}{2\pi} \ft\field(\omega_1) \int_{t_0-T}^{t_0 + T}\mathrm d t  \int_{-\infty}^{\infty} \mathrm d \tau  S_O(\tau, \omega_1) \int_{-\infty}^\infty \frac{\mathrm d \Omega}{2\pi} e^{-i\Omega(\tau - t)} \ft\field^*(\omega_1 + \Omega)\\
                           & = \frac{1}{2T} \int_{-\infty}^\infty\frac{\mathrm d \omega_1}{2\pi} \ft\field(\omega_1) \int_{t_0-T}^{t_0 + T}\mathrm d t  \int_{-\infty}^{\infty} \mathrm d \tau  S_O(\tau, \omega_1) e^{i\omega_1(t-\tau)}\field(t-\tau)
\end{align}
The $t$ integral can now be done analytically, as it includes only the
field.  The time-average effectively performs a finite-time Fourier
transform on the field,
\begin{align}
  \int_{t_0-T}^{t_0 + T}\mathrm d t   e^{i\omega_1 (t-\tau)}\field(t-\tau) =   \int_{-T}^{T}\mathrm d t   e^{i\omega_1 (t-t_0-\tau)}\field(t-t_0-\tau)
\end{align}
For an averaging interval $T$ greater than the length of the field
$t_\field$, the field can fully fit within the domain of the $t$
integral.  Hence, four regions can be defined, based on where
$\field(t-t_0-\tau)$ falls with respect to the integration limits,
\begin{align}
  \nonumber\text{ I : }& (-T-t_\field) < \tau + t_0 < (-T + t_\field)\\
  \nonumber\text{ II : }& (-T + t_\field) < \tau + t_0 < (+T-t_\field) \\
  \nonumber\text{ III : }& (+T-t_\field) < \tau + t_0 < (+T + t_\field)  \\
  \nonumber\text{ IV : }& \text{ otherwise}
\end{align}
The $t$ integrand is zero in region IV, where the $\tau+t_0$ is such
that the field is outside $[-T,T]$.  In region II, the field is
entirely contained in $[-T,T]$ and the bounds of the $t$ integral can
be extended to infinity.  A phase-insensitive term is thus obtained,
\begin{align}
 \overline O_\text{II}(t_0,2T) & = \frac{1}{2T} \int_{-\infty}^\infty\frac{\mathrm d \omega_1}{2\pi} |\ft\field(\omega_1)|^2  \int_{-T+t_\field}^{T-t_\field}\mathrm d \tau  S_O(\tau-t_0, \omega_1).
\end{align}
Finally, in regions I and III the field straddles the boundary of the
$[-T,T]$ integral,
\begin{align}
  \overline O_\text{I}(t_0,2T) & = \frac{1}{2T} \int_{-\infty}^\infty\frac{\mathrm d \omega_1}{2\pi} \ft\field(\omega_1)  \int_{-T-t_\field}^{-T +t_\field} \mathrm d \tau  S_O(\tau-t_0, \omega_1) \int_{-T}^{T}\mathrm d t e^{i\omega_1(t-\tau)}\field(t-\tau)\\
  \overline O_{\text{III}}(t_0,2T) & = \frac{1}{2T} \int_{-\infty}^\infty\frac{\mathrm d \omega_1}{2\pi} \ft\field(\omega_1)  \int_{T-t_\field}^{T +t_\field} \mathrm d \tau  S_O(\tau-t_0, \omega_1) \int_{-T}^{T}\mathrm d t e^{i\omega_1(t-\tau)}\field(t-\tau)
\end{align}
where the $t$ limits have been kept intact.  Only region I and III are
phase sensitive.

As the averaging interval becomes much larger than the field, the
phase sensitive contribution to the average disappears.  The approach
to the steady-state can be described using the scaling of phase
effects with respect to $T$ and $t_\field$,
\begin{align}
   \overline O_\text{I}(t_0,T) + \overline O_{\text{III}}(t_0,T) = \overline O_\text{coh}(t_0, T) &\propto t_\field / T\label{eq:bound1}\\
   \overline O_\text{II}(t_0,T) = \overline O_\text{inc}(t_0, T) &\propto (T-t_\field)/T
\end{align}
where the subscripts denote coherent (phase sensitive) and incoherent
(phase insensitive) contributions to the average.  For an operator
where $O(t)>0$, such as the isomer populations of
retinal,\cite{prokhorenko_coherent_2006} the coherent contribution to the
average can further be bounded.  Denoting the maximal and minimal
$O(t)$ achieved over the averaging interval $O_\text{max}(t_0, T)$ and
$O_\text{min}(t_0, T)$ after excitation with $\field(t)$, the coherent
contribution is bounded by
\begin{align}
  \overline O_\text{coh}(t_0, T) & < \frac{t_\field}{T} O_\text{max}(t_0, T)\label{eq:bound2}.
\end{align}
Then, the proportion of $\overline O(t_0, T)$ that is phase-dependent is at most,
\begin{align}
  R_c = \max \frac{\overline O_\text{coh}(t_0, T)}{\overline O(t_0, T)} = \frac{t_\field}{T} \frac{O_\text{max}(t_0, T)}{O_\text{min}(t_0, T)}.\label{eq:Rc}
\end{align}
It should be noted that this is an upper bound on the amount of phase
control --- decoherence and decay processes, for instance, further
reduce the stability of control,\cite{pachon_mechanisms_2013} while
experimental constraints on the phase manipulation of the field reduce
the available control.\cite{lavigne_interfering_2017}

\end{document}